\documentclass[11pt,twoside]{article}
\usepackage{graphicx,epsfig,natbib}
\usepackage{CS18}

\begin{document}
%
%
%
\title{Cool Stars and Space Weather}
%
%
\author{A.~A.~Vidotto$^{1}$, M.~Jardine$^{2}$, A.~C.~Cameron$^{2}$, J.~Morin$^{3}$, J.~ Villadsen$^{4}$, S.~Saar$^5$, J.~Alvarado$^6$, O.~Cohen$^5$, V.~Holzwarth$^7$, K.~Poppenhaeger $^{5,8}$, V.~R\'eville$^9$}
\affil{$^1$University of Geneva, Chemin des Mailletes 51, Versoix, 1290, Switzerland}
\affil{$^2$SUPA, School of Physics and Astronomy, University of St Andrews, North Haugh, St Andrews, KY16 9SS, UK}
\affil{$^3$LUPM-UMR5299, CNRS \& Universit\'e Montpellier II, Place E.~Bataillon, Montpellier, F-34095, France }
\affil{$^4$Department of Astronomy, California Institute of Technology, 1200 E. California Ave., Pasadena, CA 91125, USA}
\affil{$^5$Harvard-Smithsonian Center for Astrophysics, 60 Garden St., Cambridge, 02138 MA, USA}
\affil{$^6$ESO, Karl-Schwarzschild-Strasse 2, D-85748, Garching bei M\"unchen, Germany}
\affil{$^7$Kiepenheuer-Institut f\"ur Sonnenphysik, Sch\"oneckstr.\,6, 79104 Freiburg, Germany}
\affil{$^8$NASA Sagan Fellow}
\affil{$^9$Laboratoire AIM, DSM/IRFU/SAp, CEA Saclay, 91191 Gif-sur-Yvette Cedex, France}
\begin{abstract}
Stellar flares, winds and coronal mass ejections form the ``space weather\index{Space Weather}''. They are signatures of the magnetic activity of cool stars and, since activity varies with age, mass and rotation, the space weather\index{Space Weather} that extra-solar planets experience can be very different from the one encountered by the solar system planets. How do stellar activity and magnetism influence the space weather\index{Space Weather} of exoplanets orbiting main-sequence stars? How do the environments surrounding exoplanets differ from those around the planets in our own solar system? How can the detailed knowledge acquired by the solar system community be applied in exoplanetary systems?  How does space weather\index{Space Weather} affect habitability? These were questions that were addressed in  the splinter session ``Cool stars and Space Weather'', that took place on 9~Jun~2014, during the Cool Stars 18 meeting. In this paper, we present a summary of the contributions made to this session.
\end{abstract}
%
%
%
%
%
\section{Introduction}

The magnetic activity of cool stars in the form of flares, winds and coronal mass ejections (CMEs) drives the ``space weather\index{Space Weather}'' that has a direct impact on planets. This activity varies with the mass, age\index{age} and rotation rate of the star and can be damaging for life, even in the case of a fairly inactive star like the Sun. During periods of intense solar activity, the solar wind is enhanced and geomagnetic storms produce aurorae, disrupt radio transmissions, affect power grids, damage orbiting satellites, and can be hazardous for the life of astronauts. By analogy, the effects of space weather\index{Space Weather} on exoplanets may be hazardous for the creation and development of life and is therefore of potential importance for habitability\index{habitability}.

Understanding the effects of space weather\index{Space Weather} on solar system planets and exoplanets helps us not only to constrain habitability\index{habitability} but is also useful to characterise exoplanetary systems. For instance, planetary radio emission similar to that of Jupiter is expected to occur from the interaction between stellar wind particles and the planet's magnetic field \citep[e.g.,][]{1999JGR...10414025F,2001Ap&SS.277..293Z,2008A&A...490..843J,2010ApJ...720.1262V,2012MNRAS.423.3285V}. Searches are underway to detect this \citep{2000ApJ...545.1058B,2010AJ....140.1929L,2013ApJ...762...34H,2013A&A...552A..65L}, since it would characterise the planetary magnetic field strength, thus guiding studies of exoplanetary dynamos and also internal structure. The detection of transit asymmetries, which are potentially linked to the interaction of the planet and the stellar environment, can also shed light on the blow-off of atmospheres and magnetic fields\index{magnetic fields} of exoplanets \citep[e.g.,][]{2010ApJ...714L.222F,2010ApJ...721..923L,2011MNRAS.411L..46V,2012A&A...543L...4L,2013A&A...551A..63B}. Magnetic and tidal interactions between planets and host stars are expected to be revealed through observable signatures of anomalous stellar activity enhancement and spin-up \citep[e.g.,][]{2004IAUS..219..355S,2005ApJ...622.1075S,2010MNRAS.406..409F,2014A&A...565L...1P}. 

With the aim  to bring together observers/theoreticians whose diverse research interests are linked with space weather\index{Space Weather} (solar and extra-solar), Vidotto and Jardine organised the splinter session ``Cool Stars and Space Weather\index{Space Weather}'', held during  the Cool Stars 18 meeting in Flagstaff, AZ, in June 2014. In this paper, we summarise the contributions made to this session.

\section{Issues of planetary habitability\index{habitability}} 
{Andrew C. Cameron (University of St Andrews, UK)}\\

Andrew Cameron opened the session with a review talk introducing some of the issues surrounding planetary habitability\index{habitability}. Using the simplest definition, the habitable zone\index{habitable zone} of a star encloses the range of distances at which the black-body equilibrium temperature of an Earth-mass planet could be consistent with the existence of liquid surface water. The seminal paper of \citet{1993Icar..101..108K} proposes that the inner edge of the Sun's current habitable zone\index{habitable zone} lies at about $0.95$~au. Inside this distance, loss of stratospheric water by photolysis and hydrogen escape is likely to produce a runaway greenhouse. The outer edge is at about $1.37$~au, where the formation of CO$_2$ clouds raises the planetary albedo. 

The carbonate-silicate cycle and the greenhouse effect have provided an effective thermostat for most of the Earth's history. The length of time for which a planet remains continuously habitable depends on the host star's nuclear-burning lifetime. For example, in a plot of orbital separation versus stellar mass, tracks representing the inner edge of the habitable zone\index{habitable zone} at age\index{age} 5 Gyr and the outer edge at age\index{age} 700 Myr (corresponding to the earliest fossil traces of life on Earth) intersect at about $1.2~M_\odot$. The Sun is surprisingly close to the upper mass limit at which a habitable zone\index{habitable zone} can be sustained continuously up to an age\index{age} of 5 Gyr. So would life have better prospects on an Earth analogue orbiting a less massive star?

Using simple scaling laws for the stellar mass-luminosity relation, and empirical power-law relations for the decline of X-ray flux with age\index{age}, it is found that the X-ray flux in the habitable zone\index{habitable zone} at age 5 Gyr scales roughly as $M_\star^{-4}$. Using the relations of \citet{2005ApJ...628L.143W} between stellar wind mass flux and X-ray flux, the mass flux impinging on a planet in the HZ scales as $M_\star^{-4.68}L_x^{1.34}$. The solar X-ray flux has declined by a factor of 100 or so since the late heavy bombardment (LHB). Planets in the habitable zones\index{habitable zone} of stars of masses $0.32~M_\odot$ and $0.37~M_\odot$
at age 5 Gyr should experience X-ray and wind fluxes comparable with those experienced by Earth at the time of the LHB.

On planets orbiting low-mass stars, the kinetic energy of cometary impactors scales inversely with stellar mass, potentially leading to erosion of volatiles if the impact velocity exceeds twice the escape velocity $v_{\rm esc}$. Most dramatically of all, however, the amplitude of the stellar tide scales as $M_\star^2 a^{-6}$, where $a$ is the orbital semi-major axis, giving an overall dependence of $M_\star^{-10}$ for planets in the habitable zone\index{habitable zone}. \citet{1993Icar..101..108K}  point out that this should lead to locking of the planetary rotation at stellar masses less than $0.7~M_\odot$ or so.

In summary, the space-weather environments of planets in the habitable zones become progressively more extreme as stellar mass decreases. There may therefore be good anthropic reasons why the Earth orbits a star close to the maximum mass limit that allows 5 Gyr of continuous habitability\index{habitability}.

\section{Exploring a threat to foreign worlds: detecting coronal mass ejections on nearby stars} \label{sec.villadsen}
Jackie Villadsen (Caltech, Dept of Astronomy, USA)\\

The next contribution was given by Jackie Villadsen, who presented observational efforts to detect and image CMEs\index{CME} on nearby, active stars. CMEs\index{CME} likely play a significant role in the mass loss from active stars, and may significantly affect exoplanetary magnetospheres and atmospheres. On the Sun, radio flares are detected as coherent long-duration bursts and are frequently associated with CMEs\index{CME}. These radio bursts, known as Type II bursts,  trace the motion of electrons. The frequency of the emission, due to fundamental plasma radiation, is given by the plasma frequency $\nu_p = 9~{\rm kHz~} n_e^{1/2}$, where $n_e$ is the electron number density in units of cm$^{-3}$.  Type II bursts on the Sun sweep downwards in frequency on timescales of tens of minutes, tracing the motion of a CME outwards through the solar atmosphere into progressively lower plasma densities. 

Broadband dynamic spectroscopy has long been used to study coherent radio emission associated with solar CMEs\index{CME}, but such emission has not been detected from other stars. \citet{1989A&A...220L...5G} and \citet{2006ApJ...637.1016O} performed dynamic spectroscopy of  coherent radio bursts of the active M dwarf AD Leo. These bursts had short duration  and  did not show the overall slow drift in frequency that may indicate bulk plasma motion. 

Villadsen and collaborators obtained JVLA observations of the nearby star UV Ceti. They detected two coherent Type II-like radio bursts with a drift in frequency of 1.75 MHz/s. However, contrary to the solar type-II bursts, the bursts detected in UV Ceti swept upwards in frequency. They provided two possible explanations for this. First, this positive frequency drift might indicate that there is bulk plasma motion downwards in the stellar atmosphere. They estimate that this material would be falling at velocities of roughly $1000$~km~s$^{-1}$. Alternatively, the positive drift could be auroral emission from the magnetic poles (like those observed in Jupiter), which would be modulated by rotation.  

She also presented the Starburst program, a 3-year nightly observing program using two 27-meter telescopes at the Owens Valley Radio Observatory (the equivalent of a JVLA baseline).  The Starburst program will survey stellar coherent radio bursts in order to characterise the rate and energetics of CMEs\index{CME} on nearby stars, combined with complementary observations to image and characterise the detected CMEs\index{CME}.

\section{Superflares in Kepler open clusters NGC 6811 (1 Gyr-old) and NGC 6819 (2.5 Gyr-old)} \label{sec.saar}
Steve Saar (Harvard-Smithsonian CfA, USA)\\

``Superflares''\index{superflare} are a strong type of flare with energy $E>10^{33}$~erg. Their associated CMEs\index{CME} and energetic particles can damage satellites, power grids and astronauts/air travellers. They can  ionise, alter and strip atmospheres of surrounding planets; damage their ozone layers and increase mutagenic radiation, among others. They can also be an important source of losses of mass and angular momentum on young stars. 

Steve Saar presented a study on the rate of strong flares in two open clusters observed by Kepler:  NGC 6811 and NGC 6819. These clusters have estimated ages of 1~Gyr and 2.5~Gyr, respectively. The sample used was limited to single, solar-like stars with effective temperatures around solar ($\pm 200$~K). By assuming a blackbody temperature of $10^4$~K for the flares, the estimated energies of the detected flares were $ E > 3 \times 10^{33}$~erg (NGC 6811) and $E >10^{34} $~erg (NGC 6819).  

For the youngest cluster (NGC 6811),  Saar and collaborators detected $13\pm 3$ superflares\index{superflare} in a sample of 31 stars, over a time scale of 3.8 yr (i.e., a rate of $r_{1\,{\rm Gyr}} = 0.11$ flare yr$^{-1}$ star$^{-1}$). They did not detect flares on systems with hot Jupiters. For the oldest cluster,  the number of detected flares was $1\pm 1$ in a sample of 15 stars, over a time scale of 2.2 yr ($r_{1\,{\rm Gyr}} = 0.03$ flare yr$^{-1}$ star$^{-1}$). These preliminary results led Saar and collaborators to propose a relation of the decay of rates of superflares\index{superflare} with age\index{age} as  
\begin{equation} \label{eq.rate}
r \propto {\rm age}^{-1.4} \, ,
\end{equation}
consistent to the slope of $-1.5$ that one would naively expect. Using equation~(\ref{eq.rate}), they then estimated the rate of strong flares for the Sun as
\begin{equation} 
r_\odot = 0.013^{+0.03}_{-0.013} \, ,
\end{equation}
i.e., an average of one solar superflare\index{superflare} every 80~yr.

Their preliminary results indicate that younger stars have many very strong flares. At the age\index{age} of the Hyades (0.6~Gyr), they estimated a rate of flares of $0.24$~yr$^{-1}$ and a rate of $3.5$~yr$^{-1}$ at the age\index{age} of the Pleiades (0.1~Gyr), in solar-type stars. Since on the Sun most X-class flares have associated CMEs\index{CME}, stellar superflares likely will as well. This should have major implications on early habitability\index{habitability} of exoplanets and is particularly relevant for close-in exoplanets, orbiting around M dwarfs, which remain active for longer times.

\section{Cool stars magnetic fields\index{magnetic fields}}\label{sec.morin}
{Julien Morin (University of Montpellier, France)}\\

In the next talk, Julien Morin presented a brief review of cool stars' magnetic fields\index{magnetic fields}, the underlying engine of space weather phenomena. Cool stars generate their magnetic fields\index{magnetic fields} through dynamo action, hence their properties appear to correlate -- at least to some extent -- with stellar parameters such as rotation, mass and age\index{age}. We are however not yet able to completely describe these relations quantitatively under the form of scaling laws. Addressing the topic of space-weather therefore requires a twofold effort: on the one hand constraining how the properties of dynamo-generated\index{dynamo} magnetic fields\index{magnetic fields} depend on stellar parameters; and on the second hand modelling the effect of different magnetic properties on space weather phenomena (or at least on the main wind parameters, such as the angular momentum loss rate, see e.g. Section \ref{sec.reville} on Victor R\'eville's talk). It is worth noting that a better understanding of stellar dynamos is of prime interest not only for space weather issues but also to consistently address key issues of stellar physics such as the formation of low-mass stars and their planetary systems, or their rotational evolution.

Probing the magnetic fields\index{magnetic fields} of distant stars can be done through various approaches which can be broadly divided into two main categories: activity measurements and ``direct'' magnetic field measurements based on the Zeeman effect. The term ``activity'' encompasses a wide range of phenomena -- for instance spots or plage at photospheric level, chromospheres and coronae -- which can be detected across the electromagnetic spectrum from X-ray to radio wavelengths (see e.g. sections \ref{sec.villadsen}, \ref{sec.saar} and \ref{sec.poppenhaeger}). Activity proxies such as Ca~\textsc{ii} index or the ratio of X-ray to bolometric luminosity ${L_X}/{L_{\rm bol}}$ are widely used to provide a relative measurement of the average magnetic field of stars and have allowed important advances such the identification of the Rossby number as the crucial parameter for stellar dynamos (e.g.  \citealt{1984ApJ...279..763N}, \citealt{1984A&A...130..143M}, \citealt{2003A&A...397..147P}), or the detection of magnetic activity cycles on stars other than the Sun (e.g. \citealt{2007AJ....133..862H}). Informations about the field topology are however not easily derived from these measurements, although the detection of pulsed and strongly polarised radio emission \citep{2007ApJ...663L..25H,2008ApJ...684..644H} as well as the evidence for a break of the Guedel-Benz correlation between radio and X-ray luminosities (e.g. \citealt{2014ApJ...785....9W}) on cool stars and brown dwarfs open new possibilities. Direct measurements on their side rely on the Zeeman effect i.e. directly the effect of the magnetic field on the formation of spectral lines. Measurement of the Zeeman broadening in unpolarised light is a powerful tool to estimate of average of the total unsigned surface field strength irrespective of the field complexity but provides only little information about the field topology \citep[e.g.,][]{2012LRSP....9....1R}. Finally, with phase-resolved spectropolarimetric observations analysed by means of Zeeman-Doppler Imaging (ZDI\index{Zeeman-Doppler Imaging}) it is possible to reconstruct the intensity and topology of the large-scale component of stellar magnetic fields\index{magnetic fields}. But this technique tells us nothing about the small-scale field component, which is missed within the resolution element of the reconstructed ZDI\index{Zeeman-Doppler Imaging} maps. These various approaches therefore appear to be  complementary \citep[see, e.g.,][]{2013AN....334...48M}.

Since only ZDI\index{Zeeman-Doppler Imaging} is able to provide us with information about the topology of stellar magnetic fields\index{magnetic fields} and with magnetic maps which can serve as a base for space weather models, recent ZDI\index{Zeeman-Doppler Imaging} results were further discussed. During the past few years, magnetic fields\index{magnetic fields} have been detected and studied on stars throughout the Hertzsprung-Russell diagram, in particular through ambitious spectropolarimetric observing programs such as Bcool (focused on main-sequence stars), MaPP (focused on accreting T Tauri stars), MaTYSSE (on naked T Tauri stars) and TOUPIES (on zero-age and early-main sequence stars). Thanks to these observational efforts, trends in the large-scale magnetism are starting to emerge \citep{2008MNRAS.390..567M,2010MNRAS.407.2269M,2008MNRAS.390..545D,2008MNRAS.388...80P,2009ARA&A..47..333D,2012ApJ...755...97G,2014MNRAS.441.2361V}. 

Of particular interest to this splinter session are the fact that cool stars can host a wide variety of magnetic fields\index{magnetic fields}, and that even a given star can generate magnetic fields\index{magnetic fields} with very different properties throughout its evolution or even during magnetic cycles. This stresses the need for space weather models accounting for this diversity (see also Sections~\ref{sec.ofer}, \ref{sec.volkmar} and  \ref{sec.reville}). Even more central to the session is the magnetism of planet-hosting stars \citep[e.g.,][ see also Section~\ref{sec.alvarado}]{2013MNRAS.435.1451F}. Interestingly, the only star other than the Sun for which a magnetic cycle has been directly observed with spectropolarimetry is $\tau$ Boo \citep{2008MNRAS.385.1179D,2009MNRAS.398.1383F}, which has a close-in planet orbiting at 0.049~au. However, the magnetic cycle of this star has a timescale of 2~years, one order of magnitude smaller than the cycle observed in the Sun. Variability in stellar magnetism is not unique to hot-Jupiter hosts and it has been observed in several stars now \citep[e.g.,][]{2011AN....332..866M}, but such a variability is particularly more important for systems with planets orbiting close to their host stars. Stellar winds are regulated by the star's magnetism and as the magnetism varies, the conditions of the medium surrounding exoplanets also vary (\citealt{2014MNRAS.438.1162V}, see also Sections~\ref{sec.ofer} and \ref{sec.volkmar}). 

Stellar spectropolarimetry has already proven a very powerful technique to investigate cool stars' magnetism. In the next few years the availability of a new generation of instruments -- such as SPIRou, a near-infrared spectropolarimeter at CFHT -- will certainly impulse important advances. SPIRou will be capable of simultaneously measuring both the large- and small-scale fields of M dwarf stars, currently the main targets in searches of Earth-like planets orbiting in the habitable zone\index{habitable zone}. In addition, SPIRou will perform high-precision velocimetry, which will also enable it to detect planets orbiting moderately active stars. As a result, when a planet is discovered with SPIRou, the information on the host star'€™s magnetic field will be readily available.

\section{magnetic fields\index{magnetic fields} in planet-hosting G-type stars}\label{sec.alvarado}
Julian Alvarado (ESO, Garching, Germany)\\

The next contribution was from Julian Alvarado, who presented the reconstructed large scale magnetic field maps and activity diagnostics of two planet hosting Sun-like stars, HD 1237 and HD 147513, from a time-series of spectro-polarimetric data. In this work, Alvarado and collaborators have tested some of the basic assumptions behind the least-square deconvolution (LSD) techniques and confirmed the robustness of similar published maps. In this context, they refined the LSD line mask selection by applying individual line depth adjustments (tweaking) throughout the entire spectrum, achieving higher signal-to-noise ratio with the same amount of spectral lines included. 

Furthermore, with the aid of Doppler Imaging\index{Doppler Imaging} and ZDI\index{Zeeman-Doppler Imaging} techniques, they also refined the inclination angle and rotation period for both stars, essential parameters to understand the stellar magnetic field distribution and environment. At the same time, two different synthetic line shapes (Gaussian/Milne-Eddington) were tested in order to compare/improve the ZDI\index{Zeeman-Doppler Imaging} results for slowly rotating stars, such as these ones. 

The next step in this work will include these surface magnetic field maps as initial inputs for a 3D magnetohydrodynamics code (BATS-R-US), to model the large-scale topology and wind structure of the two stars. This will be used to predict the mass- and angular momentum-loss rates due to stellar winds and compare these to previous studies on more active stars. This comparison will enable further refinement of their models and get a better understanding on the physics of planet hosting-systems such as these.

\section{Magnetospheric structure and atmospheric Joule heating of planets orbiting in the habitable zone of active M-dwarfs}\label{sec.ofer}
Ofer Cohen (Harvard-Smithsonian CfA, USA)\\

Next, Ofer Cohen presented the study of the magnetospheric structure and the ionospheric Joule Heating of habitable planets orbiting M-dwarf stars using a set of magnetohydrodynamic (MHD) models. The stellar wind solution was obtained using an MHD model for the stellar corona. Cohen and collaborators then extracted the stellar wind parameters at particular locations along the planetary orbit to drive an MHD model for the planetary magnetosphere, which is coupled with a model for the planetary ionosphere. The solutions from these models provide the magnetospheric structure and the Joule heating of the upper atmosphere as a result of the interaction with the stellar wind. 

Their simulations revealed that the space environment around close-in habitable planets is extreme, with the stellar wind dynamic pressure, magnetic field, and temperature being 10 to 1000 times stronger than that at 1~au. The stellar wind plasma conditions change from sub- to super-Alfv\'enic along the planetary orbit. As a result, the magnetospheric structure changed dramatically. They note that, in a way, the transitioning between the plasma sectors mimics a CME hiting the planet. 

A significant amount of Joule heating is provided at the top of the atmosphere as a result of the planetary interaction with the stellar wind. For the steady-state solution, they found that the heating is about 0.1-3\% of the total incoming stellar irradiation, and it is enhanced by 50\% for the time-dependent case. 

This significant Joule heating should be considered in models for the atmospheres of habitable planets in terms of the thickness of atmosphere, the top-side temperature and density, the boundary conditions for the atmospheric pressure, and particle radiation and transport.

\section{Joint magnetospheres of star-planet systems}\label{sec.volkmar}
Volkmar Holzwarth (Kiepenheuer-Institut f\"ur Sonnenphysik, Freiburg, Germany) \\

The next talk was presented by Volkmar Holzwarth on joint magnetospheres of star-planet systems. The main objective of the work by Holzwarth \& Gregory was to investigate inter-connecting magnetic field structures in the joint magnetosphere of close star-exoplanet systems. They wanted to determine, on one side, possible regions of planet-induced activity signatures in the stellar atmosphere and compare them with observed phase shifts between enhanced chromospheric emission and exoplanet. On the other side, they wanted to determine planetary surface regions of inter-connecting magnetic flux which are exposed to infalling coronal material. The latter aspect is, in particular, important for planetary sciences, since high-energy particles resulting from flaring events along inter-connecting field structures in the stellar corona may interact with and heat the planetary atmosphere, modify its chemistry and change, for example, the spectral surface albedo of small, rocky planets.

\begin{figure}[!ht]
\centering
\includegraphics[width=0.5\textwidth]{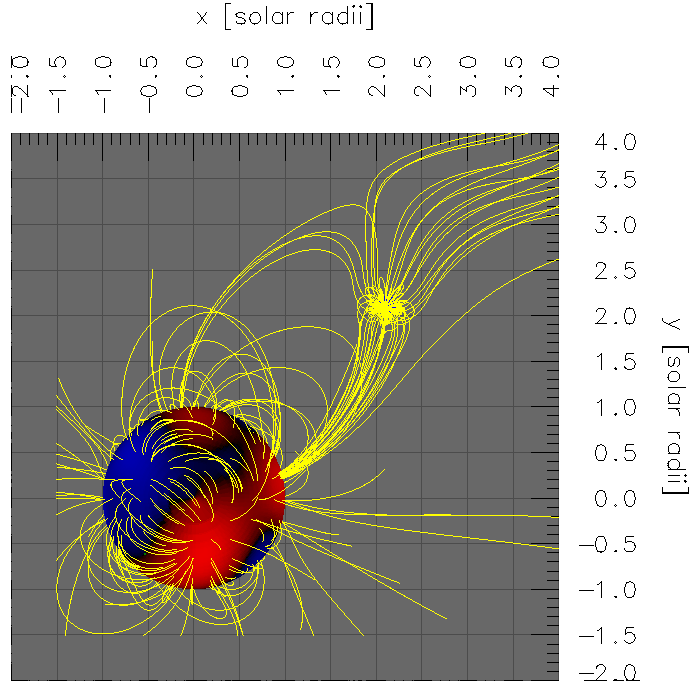}
\caption{Joint magnetosphere of the star-exoplanet system. Further details can be found in Holzwarth \& Gregory, this volume.}
\label{snap}
\end{figure}

They investigated the magnetic interaction of a hot Jupiter with its magnetically active host star in the framework of a potential magnetic field approximation (Figure~\ref{snap}). The separation of the assumed star-exoplanet system is  $3\,R_\odot$, the stellar radius $1\,R_\odot$, the radius of the Jupiter-sized exoplanet $0.1\,R_\odot$, and the source surface radius\footnote{Stellar winds are able to stretch magnetic-field lines in the radial direction. To emulate this stretching effect, beyond a radial distance known as the source surface radius, it is assumed that the stellar magnetic field is purely radial and decays with distance squared.} $4\,R_\odot$. For the host star a realistic magnetic surface map, reconstructed on the basis of Zeeman-Doppler imaging observations, was used, whereas the planetary magnetic field is taken to resemble a magnetic dipolar perpendicular to the orbital plane. The resulting joint magnetosphere shows single- or double-loop field structures inter-connecting stellar surface regions at different latitudes with opposite polarity regions around the poles of the exoplanet. Phase shifts between the stellar footpoints of these field structures and the exoplanet are up to $45^{\scriptstyle \mathrm{o}}$ and in general agreement with observational features. 

The properties of inter-connecting structures depend on the underlying magnetic field distribution on the stellar surface and on the orbital phase of the exoplanet. For instance, the type of the inter-connection changes from a single- to a double-loop during the approach and passage of magnetic polarity boundaries on the stellar surface. Changes in the planetary magnetic field hardly affect the stellar footpoints, since they are predominantly determined by the field distribution in the stellar vicinity. Any time-dependence of planetary or stellar magnetic field, such as caused by varying dynamo\index{dynamo} action, adds to the intrinsic time-dependence of the inter-connecting field structures caused by the orbiting exoplanet.

Further investigations have to be carried out to determine the time-averaged amount of planetary surface regions exposed to high-energetic stellar material. This includes the analysis of different (observed and simulated) stellar surface magnetic field distributions, for example, of slowly and rapidly rotating stars, the influence of magnetic cycles of both stellar and planetary magnetic fields\index{magnetic fields}, the contribution of higher-order magnetic moments to the planetary magnetic field, and exoplanet orbits with different inclinations and eccentricities.

%

\section{A tale of two exoplanets: the inflated atmospheres of the hot Jupiters HD 189733 b and CoRoT-2 b}\label{sec.poppenhaeger}
Katja Poppenhaeger (Harvard-Smithsonian CfA, USA)\\

Planets in close orbits around their host stars are subject to strong irradiation. High-energy irradiation, originating from the stellar corona and chromosphere, is mainly responsible for the evaporation of exoplanetary atmospheres. Katja Poppenhaeger presented multiple X-ray observations of transiting exoplanets in short orbits were conducted to determine the extent and heating of their outer planetary atmospheres. In the case of HD 189733 b, a surprisingly deep transit profile in X-rays was found, indicating an atmosphere extending out to 1.75 optical planetary radii. The X-ray opacity of those high-altitude layers points towards large densities or high metallicity \citep{2013ApJ...773...62P}. Preliminary analysis of a similar program conducted for the Hot Jupiter CoRoT-2 b indicates that the intrinsic stellar variability dominates the transit light curve. 

Recent results concerning the X-ray emission level of planet-hosting stars point towards a possible tidal influence of exoplanets on the spin and magnetic activity of their host stars. By observing pairs of wide stellar binaries in which only one component is known to host an exoplanet, it is possible to estimate the stellar age\index{age} from the activity of the planet-free star. In two cases where a strong tidal interaction of the exoplanet on the star is expected, an activity level higher than expected for the stellar age\index{age} was found. In three other systems where only weak tidal interaction is expected, the activity levels for the planet-hosting and planet-free stars are in agreement \citep{2014A&A...565L...1P}. Observations of a larger sample of such systems are in progress.

\section{The influence of magnetic topology on stellar wind braking laws} \label{sec.reville}
Victor R\'eville (CEA/Saclay, France)\\

Stellar winds are thought to be the main process in the spin-down of main-sequence stars. The extraction of angular momentum by a magnetised wind has been studied for decades, leading to several formulations for the resulting torque \citep{1988ApJ...333..236K,2012ApJ...754L..26M}. However, complex topologies of the star's magnetic fields\index{magnetic fields} have only begun to be addressed \citep[e.g.,][]{2014MNRAS.438.1162V}. 

In the last contribution of the session, Victor R\'eville presented an on-going numerical study of the effects of the field topology on the stellar wind braking law. The angular momentum-loss rate in an axi-symmetric system is $\dot{J}=\dot{M}r_A^2 \Omega$, where $\dot{M}$ is the mass-loss rate, $r_A$ is the cylindrical radius of the Alfv\'en surface and $\Omega$ is the rotation rate of the star. In a previous numerical study \citet{2012ApJ...754L..26M} found that,  for dipolar topologies of the stellar magnetic field, \mbox{$r_A/R_\star \propto (B_{\star} R_\star^2 \dot{M}^{-1} v_{\rm esc}^{-1} (1+f^2/K^2)^{-1/2} )^{m} $}, where $R_\star$, $v_{\rm esc}$ and $B_{\star}$ are the stellar radius, escape velocity and intensity of the dipolar field, respectively, $f$ is fraction of break-up spin rate, $K$ is a constant and $m=0.22$. 

Extending this study,  R\'eville and collaborators considered, in addition to a dipolar stellar magnetic field, both a quadrupolar and an octupolar field topology. They performed 60 additional stellar wind simulations, using the 2.5D, cylindrical and axisymmetric set-up computed with the PLUTO code, for various rotation rates, magnetic field strengths and escape velocities. They found that the power-law exponent $m$ varies from  $m=0.24$ for a dipolar topology to $m= 0.15$ and $m= 0.11$ for a quadrupolar and octupolar topology, respectively. They also showed that, if the braking law is given in terms of open magnetic flux $\Phi_{\rm open}$, then a unique power law, independent of the field topology, is found 
\begin{equation}\label{eq.plaw-victor}
\frac{r_A}{R_\star} \propto \left(\frac{\Phi_{\rm open}^2}{R_\star^2 \dot{M} v_{\rm esc}} {(1+f^2/K^2)^{-1/2}}\right)^{0.31} \, .
\end{equation}

Finally, by taking the axisymmetric part of a realistic magnetic topology of a young K-type star (TYC-0486-4943-1) extracted from ZDI\index{Zeeman-Doppler Imaging} maps, and of the Sun during maximum and minimum phases, Victor R\'eville showed that the values of $r_A$  extracted from their simulations follow their fit on equation (\ref{eq.plaw-victor}).

\section{Concluding remarks}
The end of the splinter session was dedicated to an open discussion that was moderated by Gaitee Hussain. The audience was actively engaged in the discussion, reflecting the increasing interest of the cool star community in the space weather\index{Space Weather} of exoplanets. Among the open issues that were brought to discussion, the lack of observational constraints on stellar winds was an important one. For instance, the stellar wind temperature profile is an essential ingredient for the acceleration of thermally-driven winds. Yet, little is known about it. What does determine the wind temperature and hence wind speed in a thermally-driven wind model? Close-in planets are convenient obstacles positioned deep inside stellar winds. What are the key observables, if any, that might allow an independent check on wind mass-loss rates and wind temperatures? 

The effects of space weather on exoplanets is a timely topic, as it has strong synergy with current and future missions and instruments, such as LOFAR, JWST, CHEOPS, SPIRou, HARPS. These missions and instruments are designed to characterise exoplanets and push further the search for terrestrial planets orbiting inside the habitable zone\index{habitable zone}. They are also aimed at characterising the host stars and their magnetism. The Kepler mission is still offering us the opportunity to study stellar activity of thousands of stars on a variety of time-scales. Thanks to collaborations such as BCool (PIs: Jeffers, Marsden, Petit), MaPP (PI: Donati) and TOUPIES (PI: Bouvier), we now have reconstructed the surface magnetic fields\index{magnetic fields} of about one hundred cool stars, paving the way for studies of space weather\index{Space Weather} along the main sequence. In the solar system, researchers studying our own Sun are benefiting from detailed observations from the Solar Dynamics Observatory, Hinode and ground-based instruments, elucidating the role of the solar activity and magnetism in the dynamics of Sun-Earth relations. All these facilities should allow for significant breakthroughs in the field of extra-solar space weather over the next decade.

\acknowledgments{
We thank the Cool Stars 18 organisers for all the prompt assistance required to run this splinter session. AAV acknowledges support from the Swiss National Science Foundation via an {\it Ambizione} Fellowship. }

\end{document}